**Research Article**

# Investigating the Impacts of Exchange Rate and Inflation on Exports: A Double Threat or Opportunity for Türkiye?

**Emre AKUSTA**
*Kırklareli University*
*emre.akusta@klu.edu.tr, ORCID: 0000-0002-6147-5443*

**Abstract**
This study analyzes the impacts of exchange rate and inflation on exports in Türkiye. Annual data for the period 1995-2023 were used in the analysis. The Johansen cointegration analysis and Dynamic Least Squares (DOLS) method were employed in the study. Identifying the cointegration relationship enabled the estimation of the long-run coefficients. The results show that an increase in the real effective exchange rate (appreciation of the Turkish lira) and inflation reduce exports with coefficients of -0.185 and -0.125, respectively. Foreign direct investment and imports, added to the study as control variables, have a positive impact on exports with coefficients of 0.117 and 0.849, respectively. These findings indicate that exchange rate stability and inflation control are priorities for improving foreign trade performance. Furthermore, policies that increase foreign direct investment and strategically manage imports complement this process.

**Döviz Kuru ve Enflasyonun İhracat Üzerindeki Etkilerinin Araştırılması: Türkiye için İkili Tehdit mi Yoksa Fırsat mı?**

**Öz**
Bu çalışma, Türkiye'de döviz kuru ve enflasyonun ihracat üzerindeki etkilerini incelemektedir. Analizde 1995–2023 dönemine ait yıllık veriler kullanılmıştır. Çalışmada Johansen eşbütünleşme analizi ile Dinamik En Küçük Kareler (DOLS) yöntemi uygulanmıştır. Eşbütünleşme ilişkisinin tespit edilmesi, uzun dönem katsayılarının tahmin edilmesine olanak sağlamıştır. Bulgular, reel efektif döviz kurundaki artışın (Türk lirasının değerlenmesi) ve enflasyonun ihracatı sırasıyla -0.185 ve -0.125 katsayılarıyla azalttığını göstermektedir. Kontrol değişkenleri olarak modele dâhil edilen yabancı doğrudan yatırımlar ve ithalat ise ihracatı sırasıyla 0.117 ve 0.849 katsayılarıyla artırmaktadır. Bu bulgular, dış ticaret performansının iyileştirilmesi açısından döviz kuru istikrarı ve enflasyon kontrolünün öncelikli politika alanları olduğunu göstermektedir. Ayrıca, yabancı doğrudan yatırım girişlerini artıran ve ithalatı stratejik biçimde yöneten politikalar bu süreci tamamlayıcı nitelik taşımaktadır.

**Anahtar kelimeler:** Döviz kuru, enflasyon, ihracat, Johansen eşbütünleşme, Dinamik OLS
**JEL Sınıflandırma Kodları:** F14, F40, F41

## 1. Introduction

The world economy has undergone a major transformation since the 1980s, when globalization gained momentum. While the economic structures of countries have become mutually interdependent, national economies have become more sensitive to external developments. Türkiye has also taken part in this global wave and experienced radical transformations in its economic structure. In this period, macroeconomic indicators such as exchange rates and inflation have become the main determinants of countries' economic performance and foreign trade dynamics. Especially export performance is constantly shaped under the influence of these two variables. Therefore, analyzing the impacts of exchange rate and inflation on exports has attracted the attention of researchers.

The exchange rate is considered one of the main determinants of foreign trade. The value of a country's national currency against foreign currencies directly affects the international competitiveness of that country. Moreover, exchange rate fluctuations especially affect export and import prices and become a determining factor in the trade balance. A depreciation of the national currency makes export prices more attractive to foreigners, while imports become more expensive. This shows that there is a dynamic relationship between the exchange rate and the balance of trade. Inflation is another macroeconomic variable closely related to the exchange rate. In an economy with high inflation, the tendency of the national currency to depreciate is stronger. This effect is reflected in export and import prices through the exchange rate and affects the trade balance (Chiu and Ren, 2019; Kyophilavong, Shahbaz, Rehman, Souksavath, and Chanthasene, 2018). The impact of the exchange rate on inflation can be explained through three main channels. The first one is the direct impact on the prices of imported goods. Changes in the exchange rate have a direct impact on inflation by changing the prices of imported goods. The second channel has an indirect impact on aggregate demand and expenditures through changes in the relative prices of tradable and non-tradable goods. The third channel is the indirect impact of the exchange rate on inflation through changes in input prices, that is, the supply channel. However, inflation also impacts on the exchange rate. High domestic inflation leads to a depreciation of the real exchange rate, weakening the national currency against foreign currencies. This can lead to an increase in exports and a decrease in imports, affecting the trade balance (Yiheyis and Musila, 2018).

Türkiye's economic history has been shaped by the dynamic relationship between exchange rates and inflation. The worldwide oil crises and economic volatility of the late 1970s led to significant deterioration in Türkiye's economic structure, and inflation became a chronic problem. High inflation disrupted domestic economic stability and weakened Türkiye's international competitiveness. During this period, various policies were implemented to combat inflation, but these policies were insufficient. In the 1980s, Türkiye abandoned its import-substitution industrialization policy and shifted to an export-led growth model. The 24 January Decisions marked the beginning of this transformation and initiated Türkiye's international expansion. Exchange rate policies were also reshaped in this period. Türkiye started to integrate into international trade networks by adopting a multiple exchange rate regime. Although the international expansion policies implemented in Türkiye after the 1980s have been an important turning point in the fight against inflation, the effects of inflation on foreign trade are still a controversial issue (Aydogan, 2004; Mangir, 2006).

The international expansion of Türkiye's economy after the 1980s has made the impact of the exchange rate on foreign trade more pronounced. Türkiye's adoption of a dirty multiple exchange rate system in 1981 allowed it to adjust the value of the national currency more flexibly according to market conditions. This was an important step in supporting exports and regulating imports. The impact of exchange rate fluctuations on exports differed in the short and long run. In the short run, depreciation of the national currency made export prices more attractive to foreign buyers but increased import costs, leading to inflationary pressures. In the long run, the stabilization of the exchange rate contributed positively to Türkiye's foreign trade performance by increasing the competitiveness of the export sector (Karakas, 2017).

The 1990s have been a period of significant economic reforms and crises for the Turkish economy. The process of international expansion that started in the 1980s gained momentum in the 1990s. In this process, Türkiye liberalized its exchange rate policies and adopted an interventionist floating exchange rate regime (exchange rate anchor). During this period, hot money inflows and speculative attacks led to significant exchange rate fluctuations. The 1994 financial crisis caused the Turkish lira to depreciate by 120 percent against the dollar. This led to a short-run increase in exports (Boratav and Yeldan, 2001). However, growth- and expenditure-oriented policies boosted domestic demand, and imports increased as the Turkish lira appreciated in real terms. This made the effects of the relationship between exchange rate and inflation on exports more complex.

In the 2000s, Türkiye's economy faced a new crisis. The financial crises of November 2000 and February 2001 necessitated a reconsideration of exchange rate policies. During this period, Türkiye completely switched to a floating exchange rate regime and allowed exchange rates to be freely determined by market conditions. This process was a critical turning point for the Turkish economy. The fully floating exchange rate regime made the export sector more sensitive to fluctuations in global markets. Exchange rate fluctuations have created both risks and opportunities for exporters. In particular, the management of exchange rate risk has gained strategic importance for exporting firms. In this period, the relationship between exchange rates and inflation became more evident. Increases in exchange rates increased inflationary pressures by raising import costs and weakening competitiveness (Ozturk and Acaravci, 2013).

The impacts of Türkiye's exchange rate and inflation dynamics on exports are a reflection of economic policies as well as developments in global markets. The 2008 global financial crisis deeply impacted Türkiye's foreign trade balance and export performance. Exchange rate fluctuations increased after the crisis, creating uncertainty for exporters. While the depreciation of the Turkish lira had a supportive impact on exports in the short run, it led to inflationary pressures as import costs rose. Although this situation may seem like an opportunity for exporters, it led to cost increases due to the import-dependent production structure and limited the competitiveness of the export sector (Yeldan, 2009; Yildirim, 2010). Therefore, analyzing the impacts of exchange rates and inflation on exports is of great importance for both evaluating the effectiveness of economic policies and developing sustainable export strategies.

In this regard, analyzing the impacts of exchange rate and inflation on exports in Türkiye is important for both economics and current policy debates. Therefore, this study analyzes the impacts of exchange rate and inflation on exports in Türkiye for the period 1995-2023. The 1995-2023 period provides an appropriate time period to analyze the interaction of the Turkish economy's foreign trade performance with exchange rates and inflation. This period is characterized by the acceleration of Türkiye's integration into the global economy, significant changes in exchange rate policies, and both single-digit and high levels of inflation. How exchange rates and inflation impact export performance during this period is a critical question for Türkiye's economic growth and competitiveness.

This study can contribute to literature in at least five ways: (1) This study is one of the most recent studies to examine the relationship between inflation, exchange rate, and exports. Analyses that consider the interaction of these three variables together are limited. This study aims to fill this gap in literature. (2) Unlike other studies, this study uses the real effective exchange rate as the exchange rate. The real effective exchange rate refers to the real value of the currency and is calculated based on the exchange basket. In this way, trade relations with different countries and inflation differentials are considered more comprehensive. (3) In addition to traditional unit root tests, unit root tests with structural breaks are also applied in econometric analyses. This approach increases the robustness of the analysis. (4) Unlike other studies, this study estimates long-run coefficients beyond cointegration analysis. This estimation provides a perspective for understanding the long-run impacts of exchange rate and inflation on exports and is critical for developing macroeconomic policy recommendations. (5) The findings obtained in the study are elaborated with econometric methods. In this way, the impacts of exchange rate and inflation policies on exports are more clearly revealed. These contributions can also guide future research and a better comprehension of the relationships between macroeconomic variables.

The rest of the paper is organized as follows: Section 2 reviews the literature, Section 3 discusses inflation, exchange rates, and export trends in Türkiye, Section 4 describes the data and methodology, Section 5 presents the empirical findings, and finally, Section 6 concludes.

## 2. Literature Review

The relationships between exchange rates, inflation, imports and exports are among the important factors shaping macroeconomic balance and trade policies. Especially in developing countries, exchange rate fluctuations can directly affect foreign trade performance, price stability, and economic growth. For this reason, the relationships between exchange rates, inflation, and foreign trade have been analyzed in the literature with different country samples and methods.

Among international studies, Poonyth and Van Zyl (2000) examined the effects of exchange rate changes on South Africa's agricultural exports and found a positive relationship between agricultural exports and exchange rates in the short and long run. Todani and Munyama (2005) analyzed the relationship between exports and exchange rate volatility in South Africa. They found a positive relationship between exports and exchange rates in both the short and long run. Analyzing the Argentine economy, Moccero and Winograd (2006) investigated the impacts of the real exchange rate on exports. The results show that while a fall in the real exchange rate increases exports to Brazil, it has a negative impact on exports worldwide. Focusing on Asian countries, Prasertnukul, Kim, and Kakinaka (2010) analyzed the relationship between exchange rate and inflation in their study covering the period 1990-2007. They found that there is no significant relationship between exchange rate and inflation. Chit and Judge (2011) investigated the impact of exchange rate volatility on exports in East Asian countries within the framework of financial sector development. The results show that exports of less financially developed economies are more adversely affected by exchange rate volatility. Focusing on the electronics sector in Malaysia, Wong and Tang (2011) find that exchange rate volatility has both short-run and long-run effects on semiconductor exports. Caglayan, Dahi, and Demir (2013) examined the impacts of real exchange rate uncertainty on manufacturing exports in emerging economies. The study finds that exchange rate uncertainty has a negative impact on trade flows. Moreover, it is emphasized that the impacts of this uncertainty may depend on the direction of trade.

Huchet and Bahmani (2013) analyzed the impacts of dollar exchange rate volatility on US agricultural exports and imports with China. They conclude that exchange rate volatility has a significant and positive long-run impact only on non-agricultural exports. They also observe that the depreciation of the dollar has an expected long-run effect on non-agricultural imports and agricultural exports. Nyeadi, Atiga, and Atogenzoya (2014) examined the impact of exchange rate on exports in Ghana and found that exchange rate has no significant impact on exports. Nyahokwe and Ncwadi (2013) also investigated the relationship between South African exports and exchange rate and found that exports are sensitive to exchange rate movements. Schaling and Kabundi (2014) analyzed the J-curve effect in South Africa. Their study indicates that there is a significant and inverse relationship between net exports and the real exchange rate in the long run. However, they also found that this relationship does not hold in the short run. Galal and Lan (2017) analyzed the relationship between inflation and foreign trade in Egypt and showed that inflation and foreign trade interact. Monfared and Akin (2017) examined the relationship between exchange rate and inflation in Iran and found that exchange rate has a direct impact on inflation. The study revealed that money supply and exchange rate positively affect inflation, but the effect of money supply is stronger than the exchange rate. Yiheyis and Musila (2018) examined the dynamics between exchange rate, inflation, and trade balance for the Ugandan economy. The results of the study show that in the long run, the depreciation of the real exchange rate increases inflation but does not have a significant impact on the trade balance.

Among the studies conducted in Türkiye, Isik, Acar, and Isik (2004) examined the long-run relationship between inflation and exchange rate for the period 1982-2003 using cointegration analysis. The results indicate that a 1% increase in the exchange rate increases inflation by approximately 0.9%. This means that the exchange rate has a significant effect on inflation. Gul and Ekinci (2006) analyzed the relationship between real exchange rate, exports, and imports for the period 1990-2006. However, no causality relationship was found from the exchange rate to imports and exports. Instead, they found a unidirectional causality from exports and imports to the exchange rate. Yilmaz and Kaya (2007) analyzed the relationship between exchange rate, imports, and exports in Türkiye for the period 1990-2004 and found that changes in the real exchange rate have no significant effect on the trade balance. However, a reciprocal causality relationship was found between imports and exports. Aktas (2010), with data for the period 1989-2008, showed that changes in the real exchange rate do not affect imports and exports, but exports are affected by changes in imports. Dincer and Kandil (2011) analyzed the asymmetric effects of exchange rate movements on export sectors. The results showed that exchange rate policy played an important role in exports during the 1996-2002 period.

Selim and Guven (2014) analyzed the relationship between exchange rate, inflation, and unemployment for the period 1990-2012. The results indicate that there is a significant relationship between the real effective exchange rate and inflation and there is a causality from exchange rate to inflation. Similarly, Turk (2016) examined the relationship between exchange rate and inflation for the period 1987-2013. The study found that the exchange rate has a significant impact on inflation, but it does not affect the exchange rate. Göçer and Gerede (2016) analyzed the relationship between imports, exports, and inflation for the period 1989-2015. The study revealed that there is a bidirectional causality relationship between inflation, exports, and imports. Bidirectional and unidirectional causality relationships were found from exports to inflation and from imports to inflation. Moreover, a long-run cointegration relationship was found between the series. Petek and Celik (2017), in their study covering the period 1990-2015, examined the relationship between exchange rate, inflation, imports and exports and found the existence of a long-run cointegrated relationship between these variables. In addition, a unidirectional causality relationship was found from exchange rate to exports and from inflation to imports.

Ayhan (2019) analyzed the impacts of exchange rate on foreign trade in the period 2005-2014. The study reveals that foreign trade revenues positively affect exports, but exports are negatively affected by the real exchange rate and exchange rate volatility. Imports, however, are positively affected by the real exchange rate and industrial production but negatively affected by exchange rate volatility. Ozer and Kutlu (2019) analyzed the relationship between exchange rate, inflation and trade balance for the period 2003-2019. The results suggest that the foreign trade balance and inflation are affected by changes in the exchange rate. However, no significant interaction was found between exchange rates, inflation and trade balance. Acaravci and Dagli (2021) investigated the impact of exchange rate variability on Türkiye's foreign trade during the period 2002-2020. They find that real exchange rate variability has a negative impact on real imports, while real exchange rate and domestic income increases have a positive impact on real imports. Finally, Aytekin and Uçan (2022) examined the relationship between exchange rate, inflation, export and import series of Türkiye between 2004 and 2019. In their study, they concluded that there is a unidirectional causality relationship from exchange rate to inflation and exports, and a bidirectional causality relationship between exports and imports. They also found that these variables interact with each other in the long run.

The literature reveals that the relationship between exchange rates, inflation and exports varies significantly both at the country level and across sectors. Studies emphasize that exchange rate variability has various impacts on trade flows and price stability, but these impacts vary depending on the structure of the economy, sectoral dynamics and policies implemented. Studies on Türkiye have examined the relationship between exchange rates, inflation, imports and exports for different periods, variables and methods. Generally, the existence of a long-run relationship between exchange rate and inflation is widely accepted in the literature. However, the relationship between exchange rates, and imports, and exports is more complex and exhibits cyclical variations. This is due to changes in Türkiye's economic structure, exchange rate regimes and macroeconomic policies over time. In particular, the impact of the exchange rate on exports changes periodically depending on global economic conditions, domestic demand and exchange rate policies. The recent inflationary pressures in Türkiye and the uncertainty in exchange rate policies suggest that the relationship between exchange rates, inflation and exports should be re-analyzed. Therefore, this study aims to analyze the impacts of exchange rate and inflation on exports in Türkiye in the light of the most recent available data.

## 3. Inflation, Exchange Rate, and Export Trends in Türkiye

The impact of exchange rates and inflation rates on exports is critical for the sustainability and competitiveness of foreign trade. Therefore, in this section, Türkiye's foreign trade performance, exchange rate, and inflation rate are analyzed graphically. The graph visually presents data on the ratio of exports to GDP, real effective exchange rate, and inflation rates between 1995 and 2023. This graph provides important information to comprehend the impacts of global economic conditions, exchange rate fluctuations, and inflation on Türkiye's export performance.

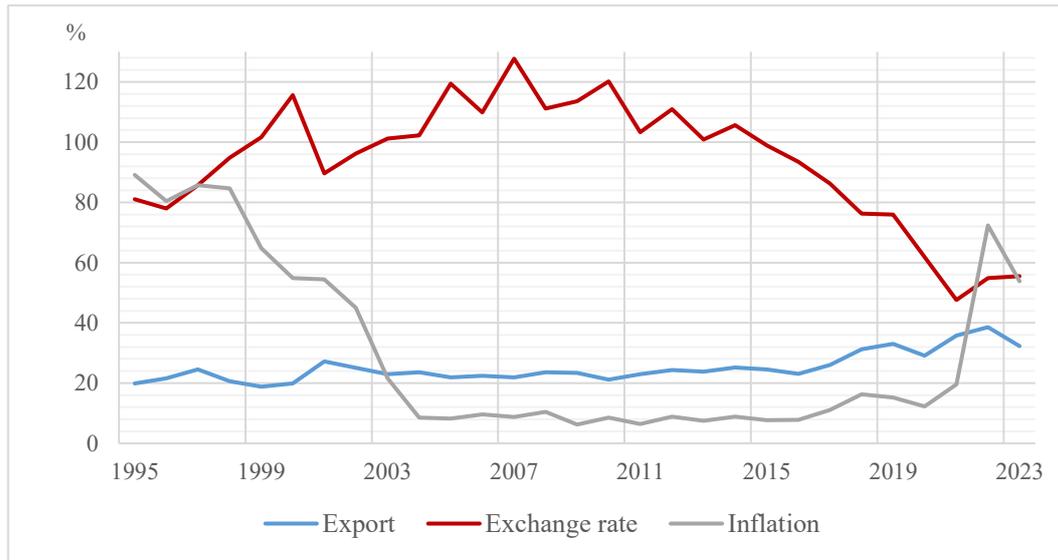

**Source: Constructed with data from CBRT (2024) and WorldBank (2024).**

**Figure 1: Selected Macroeconomic Indicators of Türkiye**

Figure 1 shows data on Türkiye's foreign trade performance, real effective exchange rate, and inflation rates for the period 1995-2023. Figure 1 shows that while exports accounted for about 19.89% of GDP in 1995, this ratio increased to 21.54% in 1996. During this period, the inflation rate fell from 89.11% to 80.41%, while the real effective exchange rate[1] declined from TL 81.03 to TL 77.98 (CBRT, 2024; World Bank, 2024). This increase in the export rate can be considered because of Türkiye's foreign trade policies and efforts to remain competitive in the global market. However, high inflation also brings economic uncertainty and rising costs. This leads to economic instability in the long run. This period was also full of uncertainties on a global scale; for example, the economic crisis in Mexico increased pressure on emerging markets and strained the economic structures of countries like Türkiye (Urus, 2019).

---

[1] A decrease in the real effective exchange rate indicates a depreciation of the domestic currency against foreign currencies.

In 1997-1998, the export ratio increased to 24.58% but dropped to 18.81% due to the impact of the 1999 Asian and Russian crises. During this period, the real effective exchange rate increased from TL 85.74 to TL 101.62, the inflation rate declined from 85.67% to 64.87%, and economic uncertainties negatively affected export performance. When the 2001 crisis is considered together with events such as the economic collapse in Argentina and the dot-com crisis in the United States, the pressure Türkiye was under during this period can be better comprehended. After the 2001 economic crisis, reforms and stabilization programs increased exports from 19.88% in 2000 to 27.18% in 2001. During this period, the real effective exchange rate decreased from 115.56 TL to 89.64 TL, and inflation decreased to 54.40% (CBRT, 2024; World Bank, 2024).

Between 2002 and 2007, Türkiye's economy experienced a period of steady growth and low inflation, with export rates stabilizing around 25%. However, the impacts of the 2008 global financial crisis reduced export performance to 23.37% in 2009, and increases in the exchange rate and inflation rates exacerbated economic difficulties. Between 2010 and 2017, export rates generally ranged between 22% and 26%, inflation remained in single digits, and the exchange rate remained relatively stable. During this period, economic policies had a positive impact on foreign trade. However, the economic downturn in 2018 pushed up the exchange rate and inflation rates. From 2018 to 2023, Türkiye's economy faced serious economic challenges. Inflation reached a historic high of 72.31% in 2022, while the real effective exchange rate increased to TL 54.88. Although the export rate reached a record high of 38.58% in 2022, high inflation and exchange rate volatility negatively affected economic stability (CBRT, 2024; World Bank, 2024).

Türkiye's economy has used various policy instruments to improve export performance and manage the impacts of exchange rates and inflation on exports. Exchange rate policies have played an important role in promoting exports and controlling imports. However, exchange rate fluctuations have increased import costs and caused inflationary pressures, adversely affecting the international competitiveness of export firms. Therefore, ensuring exchange rate stability and controlling inflation are critical for the sustainability of export performance. In conclusion, the relationship between Türkiye's foreign trade performance and exchange rate and inflation rates shows the impacts of economic policies and global economic fluctuations. However, comprehending the direct impacts of factors such as inflation and exchange rates on exports is only possible with the use of advanced econometric techniques and tools. This situation reveals the need for a detailed analysis of the relationship between inflation, exchange rates and exports in Türkiye.

## 4. Data and Methodology

### 4.1. Model Specification and Data

The empirical investigation of this study examines the impacts of exchange rate and inflation on exports. The study uses annual data for the period 1995-2023. This period was chosen based on the availability and suitability of the data set for the analysis. ADF and PP unit root tests were applied to test the stationarity of the series used in the study. In addition, the results are supported by the Zivot & Andrews unit root test that considers structural breaks. The Johansen cointegration test is used to examine the cointegration relationship between the series. Long-run coefficient estimations are performed with the DOLS method. The DOLS method is preferred because it provides more consistent and reliable estimates, especially in small samples. To assess export performance, we use exports (EXP) as the dependent variable and exchange rate (EXC), inflation (INF), foreign direct investment (FDI), and imports (IMP) as independent variables.

The model of our research is expressed in the functional form in Equation 1:

$$EXP_t = \beta_0 + \beta_1 EXC_t + \beta_2 INF_t + \beta_3 FDI_t + \beta_4 IMP_t + \epsilon_t \qquad (1)$$

In this model, $\beta_0$ represents the constant term, while the coefficients $\beta_1$ to $\beta_4$ measure the impact of each independent variable on exports. $\epsilon_t$ is the error term with zero mean and constant variance, and $t$ is the time interval. The indicators used in our study and their descriptive statistics are shown in Table 1.

### Table 1: Variable Description and Data Sources

| Variables | Symbol | Description | Obs. | Mean | SD | Min. | Max. | Source |
|---|---|---|---|---|---|---|---|---|
| Export | EXP | Exports (% of GDP) | 29 | 1.393 | 0.078 | 1.274 | 1.586 | WB |
| Exchange rate | EXC | Real effective exchange rate | 29 | 1.960 | 0.109 | 1.678 | 2.106 | EVDS |
| Inflation | INF | Consumer prices (%) | 29 | 1.286 | 0.421 | 0.796 | 1.950 | WB |
| Foreign direct investment | FDI | Net inflows (% of GDP) | 29 | 0.043 | 0.306 | -0.515 | 0.559 | WB |
| Import | IMP | Imports (% of GDP) | 29 | 1.431 | 0.076 | 1.275 | 1.629 | WB |

Note: (1) WB and EVDS indicate World Bank-World Development Indicators, and Central Bank of the Republic of Türkiye-EVDS data central, respectively. (2) The abbreviations N, SD, SE, Min and Max denote the number of observations, standard deviation, standard error, minimum and maximum values, respectively. (3) The variables used in the study are logarithmized.

Exports as a share of GDP (EXP) is an important indicator for measuring a country's foreign trade performance and its contribution to economic growth. Table 1 shows that the average export rate is 1.393% with a standard deviation of 0.078. This value indicates that exports have been relatively stable over time. The minimum value of exports is 1.274%, and the maximum value is 1.586%. The real effective exchange rate was used as the exchange rate in the study. The real effective exchange rate (EXC) is a critical factor that directly affects a country's competitiveness in international markets. The average real effective exchange rate value is 1.960 with a standard deviation of 0.109. The minimum value of this variable is 1.678, and the maximum value is 2.106. These values show that the exchange rate fluctuates over time, and it is critical for economic policy makers to analyze the effects of these fluctuations on export prices and volume. Inflation (INF), measured by the consumer price index, represents the general increase in price levels. The average inflation rate is 1.286% with a standard deviation of 0.421%. The lowest inflation rate is 0.796%, and the highest inflation rate is 1.950%. This volatility calls for an examination of the potential impact of inflationary pressures on export costs in Türkiye.

Moreover, foreign direct investments (FDI) and imports (IMP) variables are also included in this model. FDI is considered a share of net foreign investment inflows in GDP. The mean value of FDI is calculated as 0.043%, and its standard deviation is 0.306%. The minimum value is -0.515, and the maximum value is 0.559. This wide range reveals that foreign investments show significant fluctuations over time. FDI contributes to the development of new technologies and job opportunities, thereby boosting the export sector. Imports (IMP) are measured as the share of imports in GDP. The mean value of imports is 1.431% with a standard deviation of 0.076%. The minimum value is 1.275%, and the maximum value is 1.629%. Imports play a critical role in providing the raw materials and intermediate goods required for exports. An increase in imports can expose domestic producers to international competition. Therefore, analyzing the indirect impacts of imports on exports is of great importance for comprehending the balance of trade.

## 4.2. Unit Root Analysis

This study applies three different unit root tests to examine the stationarity properties of time series data: the ADF (Augmented Dickey-Fuller) test, the PP (Phillips-Perron) test, and the Zivot & Andrews structural break unit root test. These tests are used to improve the robustness of econometric models by determining whether a series is stationary or not.

*The ADF unit root test* is an extended version of the traditional Dickey-Fuller test developed by Dickey and Fuller (1979). This test addresses the problem of autocorrelation by adding lagged difference terms to the series to test for the presence of a unit root. In the ADF test, the null hypothesis ($H_0$) assumes that the series has a unit root, which means that the series is non-stationary. The alternative hypothesis ($H_1$) states that there is no unit root in the series, which means that the series is stationary. In the ADF test, if the value calculated for the test statistic is less than the critical value, the null hypothesis is rejected, and the series is considered stationary.

*The Phillips-Perron (1988) test* is an extension of the ADF test and tests essentially the same unit root hypothesis. However, the PP test considers the problems of autocorrelation and heteroskedasticity in the error terms. This test considers autocorrelation and heteroskedasticity when examining the first difference of the series. To achieve this, the Newey-West method is used. This method creates a special variance-covariance matrix to correct for autocorrelation and changing variance. The regression form of the PP test is the same as the ADF test, but the way the statistics used to correct for autocorrelation and changing variance are calculated is different. For this reason, the PP test is considered a more flexible and less parameterized method than the ADF test. The hypotheses of the PP test are like those of the ADF test: The null hypothesis ($H_0$) states that the series has a unit root, which means that the series is non-stationary. Alternative hypothesis ($H_1$) states that there is no unit root, which means that the series is stationary. The advantage of the PP test is that it eliminates some of the limitations of the ADF test by allowing correction for autocorrelation and variance.

*The Zivot & Andrews (1992) test* is a method that tests for the presence of a unit root by considering possible structural breaks in time series. Traditional unit root tests may produce misleading results since they ignore structural breaks in the series. Therefore, the Zivot & Andrews test tests for the presence of a unit root under the assumption that there may be a structural break in the series. The test has three different models: Model A considers only the break at the breakpoint; Model B considers only the break in the trend; and Model C considers both the break at the breakpoint and the trend.

### 4.3. Johansen Cointegration Test

The Johansen cointegration test is a powerful econometric tool used to identify long-run equilibrium relationships of multivariate time series. The bivariate cointegration analysis proposed by Engle and Granger (1987) was developed by Johansen (1988). Johansen's method is used to determine the number and existence of cointegration relationships in systems containing more than one series. This test is especially preferred to detect long-run relationships between variables in systems with more than one potential cointegration vector.

The main purpose of the Johansen test is to determine the rank of the $\Pi$ matrix. If the rank of this matrix is zero, there is no cointegration relationship between the variables. However, if the rank of the $\Pi$ matrix is one or more, it indicates that there is a cointegration relationship between the variables. The Johansen test provides two different statistics to determine the number of cointegration vectors: The trace Statistic and the maximum eigenvalue statistic.

The hypotheses of the Johansen test are aimed at determining the number of cointegration relationships between variables. The null hypothesis ($H_0$) states that there is a certain number of cointegration vectors between variables. The alternative hypothesis ($H_1$) suggests that there is one more than this number. The Johansen test is particularly useful for cointegration analysis of multiple series. Because it can determine whether there is more than one long-run relationship in a system. This allows for a more detailed examination of the complex long-run relationships between series and a more comprehensive analysis of the dynamic interactions between economic variables.

## 4.4. DOLS Long-run Estimates

The Dynamic Ordinary Least Squares (DOLS) method is a regression technique developed to estimate long-run coefficients in the context of cointegration analysis. Developed by Saikkonen (1991) and Stock and Watson (1993), the DOLS method aims to overcome some of the shortcomings of the traditional Ordinary Least Squares (OLS) method. It aims to obtain more consistent and reliable estimates, especially in small samples.

The DOLS method aims to eliminate the problems of autocorrelation and changing variance when determining the long-run estimation parameters. For this purpose, lagged differences of the dependent variable are added to the model. This approach reduces the bias and deviation in forecasts and provides more consistent and efficient forecasts. Moreover, the DOLS method is preferred due to its advantages, such as modeling short and long run dynamics simultaneously, minimizing endogeneity problems, and providing asymptotically consistent forecasts. For these reasons, the DOLS method is used in the analysis.

## 5. Empirical Findings

The ADF unit root test and the PP unit root test were applied to determine whether the variables used in the study contain unit roots. In addition, the Zivot & Andrews unit root test with structural breaks was also used to examine the stationarity properties of the variables by taking structural breaks into account. The Johansen cointegration test was used to analyze the long run relationship between the variables, and the DOLS method was used to estimate the long-run coefficients. The results obtained in this scope are given as follows.

**Table 2: Unit Root Test Results**

| Variable | | ADF unit root test | | PP unit root test | |
|---|---|---|---|---|---|
| | | t-statistic (level) | t-statistic (first difference) | t-statistic (level) | t-statistic (first difference) |
| EXP | Constant | 0.252 | -6.559*** | -1.495 | -8.239*** |
| EXC | | -0.486 | -6.046*** | -0.524 | -6.000*** |
| INF | | -1.456 | -4.352*** | -1.514 | -4.352*** |
| FDI | | -2.136 | -6.049*** | -1.954 | -6.789*** |
| IMP | | -1.786 | -5.491*** | -1.691 | -6.332*** |
| EXP | Constant and Trend | -2.831 | -6.888*** | -2.737 | -8.635*** |
| EXC | | -1.466 | -7.266*** | -1.466 | -9.010*** |
| INF | | -0.312 | -4.307*** | 0.814 | -5.253*** |
| FDI | | -2.491 | -5.984*** | -2.449 | -7.189*** |
| IMP | | -3.315* | -6.535*** | -3.003 | -10.956*** |

Note: The superscripts ***, **, and * denote the significance at a 1%, 5%, and 10% level, respectively.

Table 2 shows that the series are not stationary at levels based on ADF and PP test results. The null hypothesis indicating the existence of a unit root could not be rejected in either the model with a constant and the model with a constant and trend. However, since the test statistics are more negative than the critical values when the first differences of the series are taken, the null hypothesis is rejected. Therefore, it is concluded that the series do not contain unit roots. That means the series are stationary in their first differences.

**Table 3: Zivot & Andrews Structural Break Unit Root Test Results**

| Variable | Zivot & Andrews (level) | | Zivot & Andrews (first difference) | | Model |
|---|---|---|---|---|---|
| | Breakpoints | t-Statistic | Breakpoints | t-Statistic | |
| EXP | 2018 | -3.300 | 2004 | -7.252*** | Model A |
| EXP | 2015 | -3.425 | 2004 | -7.849*** | Model C |
| EXC | 2005 | -2.455 | 2001 | -7.374*** | Model A |
| EXC | 2012 | -3.289 | 2018 | -7.556*** | Model C |
| INF | 2003 | -2.538 | 2003 | -5.934*** | Model A |
| INF | 2003 | -2.851 | 2007 | -6.870*** | Model C |
| FDI | 2005 | -3.352 | 2009 | -5.720*** | Model A |
| FDI | 2005 | -3.427 | 2009 | -5.917*** | Model C |
| IMP | 2019 | -4.475 | 2009 | -8.512*** | Model A |
| IMP | 2002 | -4.875* | 2001 | -8.512*** | Model C |
| Critical value | Model A → | | %10: -4.58 | %5: -4.93 | %1: -5.34 |
| | Model C → | | %10: -4.82 | %5: -5.08 | %1: -5.57 |

Note: The superscripts ***, **, and * denote the significance at a 1%, 5%, and 10% level, respectively.

The Zivot and Andrews (1992) unit root test is applied to examine the stationarity of the series under structural breaks. The unit root test is an important analysis tool for this study, as it considers the effects of structural changes in Türkiye. In this regard, two different models of the Zivot and Andrews test are used. Model A examines the stationarity of the series by considering only the structural breaks in the constant. Model C assesses the stationarity of the series more comprehensively by considering structural breaks in both the constant and the trend.

The results in Table 3 show that in both Model A and Model C, structural breaks are detected at the level of the variables. However, the series is not stationary at this level. However, in both models, the variables become stationary in their first differences with structural breaks. In general, the results of the Zivot & Andrews unit root test with structural breaks indicate that macroeconomic variables in Türkiye have been subject to significant structural breaks over time. In particular, periods such as the 2001 economic crisis, the 2008 global financial crisis, and the 2018 currency crisis have significantly influenced the stability and trends of economic indicators such as exchange rates, inflation, foreign direct investments, and imports. These findings suggest that economic policies should be shaped not only based on domestic dynamics but also on the impacts of global economic developments. Therefore, Türkiye needs to adopt flexible and proactive policy approaches that consider the impacts of the global economic environment to ensure macroeconomic stability and sustainable economic growth.

**Table 4: Optimal Lag Length Result**

| Lag | LogL | LR | FPE | AIC | SC | HQ |
|---|---|---|---|---|---|---|
| 0 | 119.080 | NA | 1.06e-10 | -8.775 | -8.533 | -8.706 |
| 1 | 187.081 | 104.617 | 4.06e-12 | -12.083 | -10.144 | -11.665 |
| 2 | 221.472 | 39.682* | 2.51e-12* | -12.806* | -10.632* | -12.039* |

Note: Superscript * indicates the optimal lag length based on the information criterion.

Since all variables are found to be I(I), the VAR model can be constructed. However, before estimating the VAR model, the appropriate lag lengths of each system should be determined with the help of information criteria. Based on AIC (Akaike Information Criterion), SC (Schwarz Information Criterion), and HQ (Hannan-Quinn Information Criterion) information criteria, the appropriate lag length is 2 (see Table 4). Therefore, the lag length is used as 2 in this study.

## Table 5: Johansen Cointegration Test Results

| $H_0$ | Trace statistics | Critical values (5%) | Maximum eigenvalue statistics | Critical values (5%) |
|---|---|---|---|---|
| r = 0 | 75.344 | 69.819*** | 98.866 | 33.877*** |
| r ≤ 1 | 66.478 | 47.856*** | 32.638 | 27.584*** |
| r ≤ 2 | 33.840 | 29.797** | 21.686 | 21.132** |
| r ≤ 3 | 12.154 | 15.495 | 11.837 | 14.265 |
| Diagnostic tests | | | | P value |
| $\chi^2$ (Serial correlation) | | | | 0.129 |
| $\chi^2$ (Heteroskedasticity) | | | | 0.557 |
| $\chi^2$ (Normality) | | | | 0.826 |
| $\chi^2$ (Functional form) | | | | 0.248 |
| CUSUM | | | | Stable |
| CUSUMSQ | | | | Stable |

Note: The superscripts ***, **, and * denote the significance at a 1%, 5%, and 10% level, respectively.

Table 5 shows the results of the Johansen cointegration test applied to test the long-run relationship between the EXP, EXC, INF, FDI, and IMP variables analyzed in the study. This test uses trace and maximum eigenvalue statistics to determine the existence of cointegration vectors among the variables. First, under the null hypothesis r = 0 (no cointegration), the trace test statistic is 75.344, which is greater than the critical value of 69.819 at the 5% significance level. Similarly, the maximum eigenvalue test statistic of 98.866 is significantly higher than the critical value of 33.877. This data indicates that the null hypothesis is rejected and there is at least one cointegration vector. Second, when the null hypothesis r ≤ 1 is tested, the trace test statistic is 66.478 and the maximum eigenvalue statistic is 32.638, both of which are greater than the critical values. This indicates that there are two cointegration vectors. Third, under the hypothesis r ≤ 2, the trace test is 33.840 and the maximum eigenvalue is 21.686. Both exceed the critical values. This evidence supports three cointegration vectors. Finally, for the hypothesis r ≤ 3, both tests are below the critical values. This conclusion suggests that there are no more cointegration vectors.

The results show that there are at least three cointegration vectors between the variables and there is a long-run relationship between them. This implies that EXP, EXC, INF, FDI, and IMP variables move together in the long run and there is a stable equilibrium relationship between them. Furthermore, the results of the diagnostic tests to assess the validity and robustness of the model are presented in Table 5 and Table 6. The results of serial correlation, heteroskedasticity, normality, and functional form tests show that there are no statistical problems in the models. The CUSUM and CUSUM Square tests, which assess the stability of long-run forecasts, also confirm that the parameters of the model are stable over the sample period. The findings of these diagnostic tests support the statistical robustness and reliability of the analysis.

## Table 6: DOLS Long-run Estimates

| Dependent variable: EXP | Coefficient | Std. Error | t-Statistic |
|---|---|---|---|
| $EXC^2$ | -0.185*** | 0.041 | -4.512 |
| INF | -0.125*** | 0.027 | -4.654 |
| FDI | 0.117** | 0.046 | 2.569 |
| IMP | 0.849*** | 0.065 | 3.070 |
| C | 2.073** | 0.984 | 2.107 |
| Diagnostic tests | | | P value |
| $\chi^2$ (Serial correlation) | | | 0.886 |
| $\chi^2$ (Heteroskedasticity) | | | 0.295 |
| $\chi^2$ (Normality) | | | 0.536 |
| $\chi^2$ (Functional form) | | | 0.236 |
| CUSUM | | | Stable |
| CUSUMSQ | | | Stable |

Note: The superscripts ***, **, and * denote the significance at a 1%, 5%, and 10% level, respectively.

Table 6 contains the results of DOLS long-run estimates. The results show that the coefficient of EXC is -0.185 and statistically significant at the 1% level. This indicates that an increase in the real effective exchange rate (appreciation of the Turkish lira) has a negative impact on exports. This leads to a decrease in economic competitiveness. In developing countries such as Türkiye, the impact of exchange rate on exports is significant. An appreciation of the Turkish lira makes Türkiye's products more expensive abroad. This can lead to lower exports in Türkiye's main export markets, especially in highly competitive regions such as the European Union. In addition, high exchange rate fluctuations could make it harder for businesses to manage their costs and affect investment plans. In the long run, the appreciation of the Turkish lira could reduce exporters' profit margins and increase the attractiveness of imports, leading to a widening trade deficit. This could put pressure on economic growth and foreign exchange reserves. For Türkiye to sustain its export performance, it is critical to manage exchange rate policies in a stable manner and redirect exports towards high value-added products.

The coefficient of INF is -0.125 and is statistically significant at the 1% level. This result indicates that an increase in inflation has a negative impact on exports. Inflation increases production costs and weakens the competitiveness of domestic goods in international markets. In an import-oriented country like Türkiye, an increase in inflation both increases input costs and raises the prices of exported goods. This leads to a decline in exports, especially in the textile, automotive supply industry, and food sectors, where price competition is intense. An uncontrolled increase in inflation leads to uncertainty in the economy, making it difficult for exporters to forecast costs. This uncertainty may prevent exporters from entering into long-run contracts and expanding into new markets. During periods of high inflation in Türkiye, exporters lose their competitive advantage, while a contraction in domestic demand may also adversely affect exporters. From an economic perspective, keeping inflation stable and low is crucial for the sustainability of exports and the overall stability of the economy. Such an outcome is essential for Türkiye to achieve its economic growth targets and become more competitive in global trade.

---

[2] EXC represents the real effective exchange rate. An increase in the real effective exchange rate indicates an appreciation of the domestic currency against foreign currencies. The reverse is also correct.

The coefficient of FDI is 0.117 and is statistically significant at the 5% level. This indicates that an increase in net FDI inflows has a positive impact on the percentage of exports. FDI increases production capacity in Türkiye and supports exports by facilitating technology transfer, knowledge accumulation, and access to international markets. Türkiye is considered as an attractive market for foreign investors due to its geographical location and large domestic market. Increased FDI can stimulate the production of high value-added products. For example, foreign investments in the automotive and electronics sectors can strengthen Türkiye's integration into the global value chain by increasing export capacity in these sectors. Production plants established by foreign investors in Türkiye improve domestic supply chains and facilitate exporting firms' access to global markets. In economic terms, the positive impact of FDI on exports contributes to Türkiye's sustainable growth. However, for these investments to be sustainable, economic and political stability must be maintained, the legal system must be transparent, and investor rights must be guaranteed.

The coefficient of IMP is 0.849, which is positive and statistically significant at the 1% level. This result shows that imports have a significant and positive impact on exports. Especially in a country like Türkiye, where the use of imported raw materials and intermediate goods is high in the production structure, an increase in imports may lead to an increase in exports. A large portion of Türkiye's exports is import-oriented. This is especially valid in sectors such as automotive, machinery, and electronics. In these sectors, imports are vital for sustaining and increasing production. However, an import-dependent production structure creates vulnerability to exchange rate fluctuations and can widen the foreign trade deficit. While it is positive from an economic perspective that imports support exports, it is important to balance this dependence. Türkiye's import dependence negatively affects foreign exchange reserves and increases economic vulnerability to uncertainties in global markets. To sustain Türkiye's exports in the long run, Türkiye needs to reduce its dependence on imports by implementing policies that support domestic production. This will contribute to both improving the foreign trade balance and ensuring economic stability.

The findings of the study show that the real effective exchange rate and inflation have a negative impact on exports in Türkiye, while foreign direct investment and imports play a supportive role for exports. These results suggest that Türkiye should focus on policies such as exchange rate stability, inflation control, and incentives for foreign investors to increase its competitiveness in global markets. The findings are consistent with Selim and Guven (2014), Turk (2016), Gocer and Gerede (2016), Petek and Celik (2017), Ozer and Kutlu (2019), and Aytekin and Uçan (2022), who find a cointegration relationship between exchange rate, inflation, and exports. Moreover, the long-run estimates of this study are consistent with the findings of Ozer and Kutlu (2019) and Aytekin and Uçan (2022). These studies also find that depreciation of the Turkish lira supports exports, while inflation reduces exports.

## 6. Conclusion

The impacts of exchange rate and inflation on Türkiye's export performance are important for both economics and current policy debates. Therefore, this study analyzes the impacts of exchange rate and inflation on Türkiye's exports using data for the period 1995-2023. ADF and PP unit root tests were applied to test the stationarity of the series used in the study. In addition, the results are supported by the Zivot & Andrews unit root test that considers structural breaks. The Johansen cointegration test is used to examine the cointegration relationship between the series. Long-run coefficient estimations are performed with the DOLS method.

The Johansen cointegration results show that there are at least three cointegration vectors among export, exchange rate, inflation, foreign direct investment and import variables. This reveals the existence of a long-run and stable equilibrium relationship between these variables. The identification of the cointegration relationship enabled the estimation of the long-run coefficients. The DOLS findings indicate that the appreciation of the Turkish lira has a statistically significant and negative impact on exports in the long run. As Turkish goods and services become more expensive in foreign markets, competitiveness decreases and exports fall. Moreover, high exchange rate fluctuations make it difficult for businesses to manage their costs and negatively impact their investment plans. Therefore, stable management of exchange rate policies is vital for sustainability of export performance and enhancement of competitiveness. Similarly, an increase in inflation rates is also found to have a negative impact on exports. High inflation increases production costs and weakens the competitiveness of domestic products in international markets. This impact is more pronounced in the textile, automotive supply industry and food sectors where price competition is intense. Uncontrolled inflation increases economic uncertainty, making it difficult for exporters to forecast costs and plan for the long run. Therefore, keeping inflation stable and low is critical for export sustainability and overall economic stability.

The study also found that FDI has a positive impact on exports. Increased FDI contributes to export growth by increasing production capacity, facilitating technology transfer and supporting access to international markets. However, for this positive impact to be sustainable, economic and political stability, transparency of the legal system and protection of investors' rights are required. The positive impact of imports on exports is also among the important findings of the study. Due to Türkiye's import-oriented production structure, imports of raw materials and intermediate goods play an important role in increasing exports. However, this dependence on imports makes the economy vulnerable to exchange rate fluctuations and leads to a widening foreign trade deficit. Therefore, implementing policies that support domestic production and reduce import dependency is essential for sustainable exports.

The analysis reveals that the real effective exchange rate and inflation are dual threats to Türkiye's export performance. Therefore, a series of policy recommendations have been developed to overcome these threats and enhance Türkiye's export capacity. Firstly, the Central Bank of the Republic of Türkiye (CBRT) should adopt a proactive exchange rate policy to reduce exchange rate fluctuations and keep the exchange rate stable. This would contribute to protecting exporters against exchange rate uncertainty and improving cost estimates. CBRT should use various instruments to strengthen foreign exchange reserves and intervene in foreign exchange markets when necessary. Inflation is another critical issue. Implement contractionary monetary policies to reduce inflation and keep it at a stable level. CBRT should ensure price stability by controlling the money supply within the framework of inflation targeting. Moreover, maintaining fiscal discipline and reducing budget deficits also help to reduce inflationary pressures. In addition, Türkiye's investment climate should be improved to attract foreign direct investment. Legal regulations and property rights should be guaranteed, bureaucratic barriers should be reduced, and investment incentives should be expanded. Long-term investments by foreign investors will support exports through technology transfer and increased production capacity. Another recommendation is to promote the production of high value-added products to reduce import dependency and increase the sustainability of exports. Türkiye should increase R&D investments, focus on innovative production techniques and invest in high-tech sectors. This approach would help Türkiye diversify its export structure and make it more resilient to the adverse impacts of exchange rates and inflation. Moreover, strategic import policies should be developed, considering that imports support exports. In particular, imports of raw materials and intermediate goods used in exports should be cost-effective and domestic supply chains should be developed. This could alleviate pressure on the exchange rate by reducing the demand for foreign exchange. Another important step is the introduction of financial instruments and support mechanisms to protect exporters against exchange rates and inflation risks. In this regard, the availability of export insurance, credit guarantee funds and hedging instruments to reduce FX risk should be increased. Such support would enhance the competitiveness of exporters in global markets. It is also a crucial step for Türkiye to diversify its export markets and explore new ones. Expanding free trade agreements and developing specific strategies for target markets can create an export structure that is more resilient to external demand shocks. Finally, Türkiye needs to invest in its logistics infrastructure to increase its export capacity. Ports, rail and road networks should be improved to ensure that export products reach international markets quickly and cost-effectively. This will enable Türkiye to play a more effective role in global supply chains. Moreover, a qualified labor force is necessary to ensure the sustainability of high value-added production and exports. Restructuring the education system and vocational training programs in line with the needs of export-oriented sectors will increase productivity and strengthen Türkiye's global competitiveness.

Although this study provides important findings, it has some limitations. Future studies can address these limitations. First, only Türkiye was selected as the sample in this study. Future studies can expand the sample to include both developed and developing countries. In this way, it can be compared whether the results differ across countries with different income levels. Second, this analysis is limited to the period 1995-2023. Future studies may use larger data sets to examine the effects of periodic changes in more detail. Finally, larger data sets and different analysis techniques can be used to test the consistency of the findings of this study.